\documentclass[%
 reprint,
 amsmath,amssymb,
 aip,
]{revtex4-2}

\usepackage{graphicx}
\usepackage{dcolumn}
\usepackage{bm}
\usepackage{hyperref}
\usepackage{physics}
\usepackage{siunitx}

\usepackage[dvipsnames]{xcolor}

\usepackage{soul}
\usepackage{bm}

\renewcommand{\vec}{\bm}

\begin{document}

\title{Intermolecular interactions in optical cavities: an ab initio QED study}
\author{Tor S. Haugland}
\affiliation{Department of Chemistry, Norwegian University of Science and Technology, 7491 Trondheim, Norway}
\author{Christian Sch\"afer}
\affiliation{Max Planck Institute for the Structure and Dynamics of Matter and Center Free-Electron Laser Science, Luruper Chaussee 149, 22761 Hamburg, Germany}
\author{Enrico Ronca}
\affiliation{Istituto per i Processi Chimico Fisici del CNR (IPCF-CNR), Via G. Moruzzi, 1, 56124, Pisa, Italy}
\affiliation{Max Planck Institute for the Structure and Dynamics of Matter and Center Free-Electron Laser Science, Luruper Chaussee 149, 22761 Hamburg, Germany}
\author{Angel Rubio}
\affiliation{%
Max Planck Institute for the Structure and Dynamics of Matter and Center Free-Electron Laser Science, Luruper Chaussee 149, 22761 Hamburg, Germany}%
\affiliation{%
Center for Computational Quantum Physics (CCQ), The Flatiron Institute, 162 Fifth avenue, New York NY 10010.}%
\affiliation{%
Nano-Bio Spectroscopy Group, Departamento de F{\'i}sica de Materiales, Universidad del Pa{\'i}s Vasco, 20018 San Sebastian, Spain}%
\author{Henrik Koch}
\email{henrik.koch@sns.it}
\affiliation{Scuola Normale Superiore, Piazza dei Cavalieri, 7, 56124 Pisa, Italy}
\affiliation{Department of Chemistry, Norwegian University of Science and Technology, 7491 Trondheim, Norway}

\date{\today}

\begin{abstract}
Intermolecular bonds are weak compared to covalent bonds, but they are strong enough to influence the properties of large molecular systems.
In this work, we investigate how strong light-matter coupling inside an optical cavity can modify these intermolecular forces.
We perform a detailed comparison between currently available \textit{ab initio} electron-photon methodologies.
The electromagnetic field inside the cavity can modulate the ground state properties of weakly bound complexes.
Controlling the field polarization, the interactions can be stabilized or destabilized, and electron densities, dipole moments, and polarizabilities can be altered.
We demonstrate that electron-photon correlation is fundamental to describe intermolecular interactions in strong light-matter coupling.
This work proposes optical cavities as a novel tool to manipulate and control ground state properties, solvent effects, and intermolecular interactions for molecules and materials.
\end{abstract}

\maketitle

\section{Introduction}

\begin{figure*}[ht!]
    \centering
    \includegraphics[width=\textwidth]{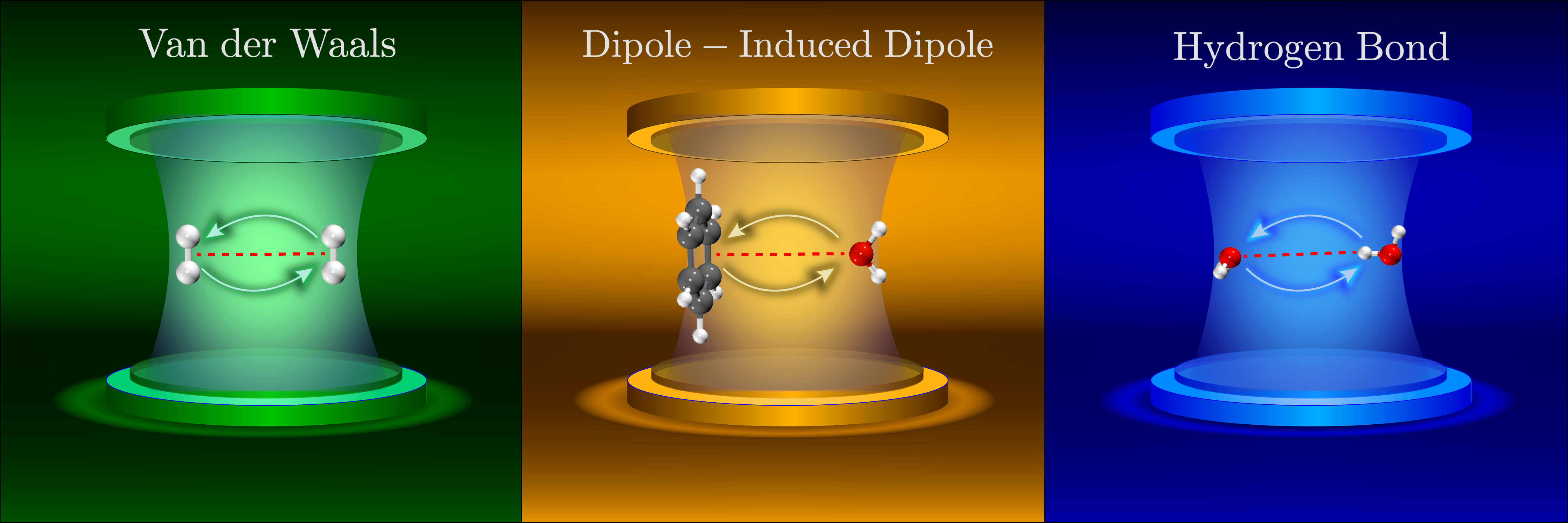}
    \caption{Main intermolecular interactions, here in optical cavities.}
    \label{fig:cartoon}
\end{figure*}

Intermolecular interactions play a fundamental role in chemistry and physics, and are especially important in describing the properties of large systems.
In particular, they contribute to solvation processes,\cite{atkins2016physical} interactions in liquids,\cite{Heald1962solvent} gas phase reactivity,\cite{Garver2010gasphase} formation of higher order structures in biological macromolecules,\cite{Lehninger2012} and multi-layer 2D materials.\cite{Cao2018moire,Kerelsky2019moire,hill2020moire,Xian2019bn}
The ability to induce even minor modifications in the intermolecular forces can have large impact on the macroscopic properties of molecular systems.
In this respect, strong light-matter coupling is an exciting possibility for changing weak interactions.

Over the last decade, strong light-matter coupling via optical cavities has been unveiled as a new tool that can modify molecular properties and interactions in a non-intrusive fashion.
Recent seminal experimental works have demonstrated
the possibility to inhibit,\cite{Hutchison2012,thomas2016ground,munkhbat2018suppression} steer,\cite{ThomasScience2019} and accelerate\cite{HiuraChemrxiv2018,Lather2019,herrera2016,campos2019resonant} chemical reactions by strong light-matter coupling inside micro\cite{Agranovich2003,wang2017coherent} and plasmonic\cite{Ojambati2019,wersall2017observation} cavities.
Furthermore, cavities have been applied to: enhance charge and energy transfer,\cite{orgiu2015,zhong2017energy,hagenmuller2017cavity,saez2018organic,SchaeferPNAS2019,groenhof2019tracking,hertzog2019strong} design materials,\cite{huebener2020} and control superconductivity. \cite{thomas2019exploringsuperconductivity,sentef2018,schlawin2019cavity}
These applications highlight the use of confined light in chemistry.

From a theoretical point of view, strong light-matter coupling has been extensively investigated since the 1950s using a simplified descriptions of the electronic system.\cite{dicke1954,jaynes1963,cummings1965}
Using model Hamiltonians, a broad range of fascinating insight has been achieved. However, connecting these predictions to experimental observations in polaritonic chemistry is quite challenging.\cite{anoopcomment,galego2019cavity,herrera2020photochemistry,li2020resonance,vurgaftman2020negligible}
The models typically account for the main features of strong light-matter coupling, such as Rabi splittings,\cite{shore1993jcm} but are unable to quantitatively capture changes in molecular systems.
In situations where the coupling between light and matter is of the same magnitude as other interactions, for instance intermolecular forces, the simplified models are not sufficient.
In order to achieve accuracy in chemical predictions, the full complexity of the system must be considered. 
\textit{Ab initio} methodologies, such as quantum electrodynamical density functional theory (QEDFT)\cite{RuggenthalerPhysRevA2014} and quantum electrodynamics coupled cluster theory (QED-CC)\cite{haugland2020coupled}, provide an accurate modeling of the correlated electron-photon systems.

In this paper, we employ the QEDFT and QED-CC methods to study intermolecular forces under the influence of strong light-matter coupling.
We focus on three selected systems representative of: van der Waals interaction, dipole-induced dipole interaction, and hydrogen bonding (see Figure \ref{fig:cartoon}).
We also benchmark, where possible, against the exact QED full configuration interaction (QED-FCI) method. 
For the hydrogen bonded system, we also analyze changes in the electron density.
Finally, we investigate the long-range intermolecular interaction through the cavity field.
We note that all the observed effects are for the ground state of the light-matter system.

The paper is organized as follows.
In the first section, we introduce the computational approaches.
The following section presents our investigation of cavity induced effects on intermolecular interactions.
The final section contains our concluding remarks.

\section{Computational Approaches}
In nonrelativistic QED, interactions between molecules and quantized electromagnetic fields are described by the Pauli-Fierz Hamiltonian.\cite{SpohnBook2004,cohen1997photons} For a single photon mode in the dipole approximation, the time-independent Hamiltonian in Coulomb gauge and after subsequent Power-Zienau-Woolley transformation\cite{power1959coulomb,schafer2019relevance} reads
\begin{equation}
\begin{split}
\label{eq:hamiltonian_elec}
    H_e =& -\frac{1}{2} \sum_i^{N_e} \nabla_i^2 - \sum_i^{N_e} \sum_I^{N_{\text{nuc}}} \frac{Z_I}{|\vec{r}_{i} - \vec{R}_{I}|}
    \\
    &+ \frac{1}{2} \sum_{i\neq j}^{N_e} \frac{1}{|\vec{r}_i - \vec{r}_j|} + \frac{1}{2} \sum^{N_{\text{nuc}}}_{I\neq J} \frac{Z_I Z_J}{|\vec{R}_I - \vec{R}_J|}
\end{split}
\end{equation}
\begin{equation}
    \label{eq:hamiltonian}
    H = H_e + \omega b^\dagger b + \lambda\sqrt{\frac{\omega}{2}} (\vec{\varepsilon} \cdot \Delta \vec{d}) (b+b^\dagger)
    + \frac{\lambda^2}{2} (\vec{\varepsilon} \cdot \Delta \vec{d}\,)^2.
\end{equation}
Here $H_e$ is the electronic Hamiltonian in the Born-Oppenheimer approximation in the standard notation.
The second term of Eq.~\eqref{eq:hamiltonian} is the photon contribution to the energy where $\omega$ is the photon mode frequency $\omega$ and $b/b^\dagger$ are the associated annihilation/creation operators.
The third term is the bilinear light-matter interaction, and contains the fluctuations of the molecular dipole, $\Delta\vec{d} = \vec{d} - \expval{\vec{d}}$, and the field transversal polarization vector $\vec{\varepsilon}$ with a coupling strength determined by the quantization volume $V$, $\lambda=\sqrt{4\pi/V}$.
The last term is the dipole self-energy (DSE) that describes the molecular self-interaction through the field and further ensures that the total Hamiltonian is gauge and origin invariant, and bound from below.\cite{schafer2019relevance,haugland2020coupled}
The same Hamiltonian in second quantization can be found in Ref.~\onlinecite{haugland2020coupled}.

In quantum chemistry, there are two main approaches to solve the electronic Schr{\"o}dinger equation:
one is based on wave functions,\cite{HelgakerBook2000} the other on one-electron densities.\cite{ParrBook1989}
To solve the eigenvalue problem for the light-matter Hamiltonian [Eq. \eqref{eq:hamiltonian}], photons must be treated on equal footing as the electrons.
Considering that photons are inherently different from electrons because of the bosonic nature, they show different physics.
For this reason, at the moment only a few molecular approaches have been extended to quantum electrodynamics.
In particular, the extensions of DFT, Hartree-Fock, CC, and FCI are now available to perform \textit{ab initio} simulations of molecules strongly coupled to light.
In the following subsections we give a brief introduction to these methodologies.

\subsection{QED wave function approaches}

Hartree-Fock theory (HF), which employ a single Slater determinant, is the standard starting point for developing approximate wave function methods.
This approach does not include any electron correlation, which is instead recovered by post-HF methodologies.
One of the most successful and accurate methods to include correlation is CC, which includes additional determinants through an exponential parameterization starting from the HF wave function.\cite{HelgakerBook2000}

In Hartree-Fock theory, the photonic degrees of freedom are included by treating the electrons and photons as uncorrelated particles interacting through a mean-field potential. This leads to the formulation of quantum electrodynamics Hartree-Fock theory (QED-HF).\cite{haugland2020coupled}
Similarly to coupled cluster theory for electrons, QED-HF is the starting point for QED-CC.\cite{haugland2020coupled}
The correlated QED-CC wave function is expressed as
\begin{equation}
    \ket{\text{CC}} = \exp(T) \ket{R}
\end{equation}
where $\ket{R}=\ket{\text{HF}}\otimes \ket{0}$ is the QED-HF state. Here $T$ is the cluster operator,
\begin{equation}
    T = \sum_{\mu n} t^{n}_{\mu} \tau_{\mu} (b^\dagger)^n
\end{equation}
and $t_{\mu n}$ are the amplitudes for the determinant $\ket{\mu}$ and photon occupation $\ket{n}$, and $\tau_{\mu} (b^\dagger)^n$ is the corresponding excitation operator,
\begin{equation}
    \tau_\mu (b^\dagger)^n \ket{R} = \ket{\mu} \otimes \ket{n} \sqrt{(n+1)!}.
\end{equation}

The QED-CC formulation is exact in the limit where all electronic and photonic states are included.
In practice, because of the infinite number of states, a truncation must be introduced.
In this paper, we consider QED-CCSD-1, which employ a cluster operator
\begin{equation}
    T = T_1 + T_2 + S^{1}_{1} + S^{1}_{2} + \Gamma^1,
\end{equation}
where $T_1$ and $T_2$ are linear combinations of electronic single and double excitations, $S^{1}_{1}$ and $S^{1}_{2}$ are electronic single and double excitations coupled with a single photon creation, and $\Gamma^1$ is the single photon creation operator.
The QED-CCSD-1 scales as $N^6$ with respect to the number of electronic basis functions, and it is computationally feasible for small to medium sized molecules.

For a fixed number of photons, if all electronic configurations are included, QED-CC is equivalent to the exact treatment of QED-FCI.
However, implementing QED-FCI in a coupled cluster formulation is inconvenient.
For this reason, we use a direct CI implementation\cite{HelgakerBook2000} of QED-FCI as a benchmark.

All the mentioned wave function based methods are implemented and calculated using the $e^T$-program.\cite{eTpaper}

\subsection{QED density functional theory}

Quantum electrodynamical density functional theory is based on the key observation of ordinary DFT that all observables relevant in electronic structure theory are functionals of the electronic density $n(\textbf{r})$. Leveraging the unique relation between electronic density and external potential,\cite{hohenberg1964} DFT allows to set up and computationally solve a set of single-particle Kohn-Sham equations.\cite{kohn1965} The occupied orbitals $i$ fulfill
\begin{align}
    \left[-\frac{1}{2} \nabla^2 + v_s(\vec{r}) \right] \varphi_i(\vec{r}) = \epsilon_i \varphi_i(\vec{r}) 
\end{align}
and provide the exact electronic density 
\begin{equation}
   n(\textbf{r}) = \sum_{i} \vert \varphi_i(\vec{r}) \vert^2
\end{equation}
associated to the potential 
\begin{equation}
    v_{s}(\vec{r}) = v_{ext}(\vec{r}) + v_{H}(\vec{r}) + v_{xc}(\vec{r}).
\end{equation}
This local Kohn-Sham potential is characterized by the known external and classical Hartree potentials as well as the unknown exchange-correlation potential $v_{xc}(\vec{r})$. Naturally $v_{xc}(\vec{r})$ is known only approximately and DFT is (in practice) often accurate and computationally efficient but never exact. 

The QEDFT approach extends this idea to quantum electrodynamics.\cite{RuggenthalerPhysRevA2014} Through the formulation of effective single-particle Kohn-Sham and Maxwell equations, QEDFT reproduces the exact many body problem if the exact functionals for the electron-electron and electron-photon interactions are known.
In this paper, we utilize the optimized effective potential (OEP) approach \cite{kuemmel2003b,pellegrini2015,FlickACSPhoton2018} to obtain the exact-exchange potential 
\begin{align}
    v_{xc}(\textbf{r})  \approx v_{x}^{\text{OEP}, \text{Coulomb}}(\textbf{r})  + v_x^{\text{OEP}, \text{photons}}(\textbf{r})
\end{align}
for both the Coulomb and photonic interaction. The latter is the variational derivative
\begin{equation}
   v_x^{\text{OEP}, \text{photons}}(\textbf{r}) = \delta E_x^{\text{photons}}/\delta n(\textbf{r})
\end{equation}
of the self-consistent second-order correction to the energy \cite{pellegrini2015,FlickACSPhoton2018}
\begin{align}
    E_x^{\text{photons}} = \frac{1}{2}\lambda^2 \sum_{a,i} \vert \langle \varphi_a \vert \vec{\varepsilon} \cdot \vec{d} \vert \varphi_i \rangle \vert ^2 \left( 1 - \frac{\omega}{\epsilon_a-\epsilon_i + \omega} \right)
\end{align}
where $a$ labels the unoccupied Kohn-Sham orbitals. Using density-functional perturbation theory, the density dependence of the Kohn-Sham orbitals lead to coupled Sternheimer (coupled perturbed Kohn-Sham) equations that are solved self-consistently.\cite{FlickACSPhoton2018,tancogne2019octopus}

The combination of QEDFT and the OEP functional (QEDFT/OEP) gives rise to an exact-exchange treatment of the electronic structure in the presence of the cavity. In this way, it describes the interaction of electrons on similar but not equivalent level as Hartree-Fock.\cite{kuemmel2003b}
In contrast to QED-HF, the QEDFT/OEP solution features also electronic exchange mediated by the photonic fluctuations in addition to the exchange originating from the DSE.
For the systems investigated here, the correct long-range $1/r$ decay is an especially desirable feature of the OEP that is not common to all possible realizations of DFT.
QEDFT itself is a much more versatile concept than the here selected OEP approach and correlation functionals are currently in development.

All reported calculations have been performed with the OCTOPUS code\cite{tancogne2019octopus} using the standard Troullier-Martins pseudopotentials, a Cartesian grid and 8th-order finite differences.

\section{Results}

Covalent bonds between valence electrons typically have binding energies of about $1$-$10$ eV.
Intermolecular energies are substantially weaker ($1$-$300$ meV) and they arise from van der Waals interaction
or electrostatics (dipole-dipole, etc.).\cite{atkins2011molecular}
Even though their strength is orders of magnitudes weaker than covalent and ionic bonds, they are not negligible.
Intermolecular forces can also be interpreted in terms of electron interactions mediated by the transverse electromagnetic field.\cite{CraigBook1984}
Because of this parallelism, changing the boundary conditions of the field with a cavity will alter the interactions.\cite{buhmann2013dispersionI,buhmann2013dispersionII,salam2016non}

In this paper, we treat a single cavity mode interacting explicitly with the molecular system, and assume that the electron mass inside the cavity is the same as outside.\cite{schaefer2018bridging,CraigBook1984} 
For QED-HF, additional cavity modes would increase the cavity induced effects on the ground state compared to a single-mode approximation.
To simulate potential multi-mode effects, we instead enlarge the coupling value to $\lambda=0.1$.
The description is certainly limited and going beyond this will require more sophisticated considerations from the photonic point of view.
However, new insight into a non-perturbative description of cavity mediated weak interactions can be obtained using this approximation.
Cavity losses, which only leads to quantitative and not qualitative changes of the results, are neglected in this paper.\cite{liberato2017virtual}

In the following, cavity induced effects on the different kinds of intermolecular interactions are discussed.  
These cases range from weak to strong intermolecular forces.

\subsection{Van der Waals interaction}

The weakest among the intermolecular interactions is van der Waals (vdW),\cite{atkins2011molecular} characterized by typical energies of about 1-50 meV.
Usually, they are described as interactions between the polarizabilities of the constituent fragments $A$ and $B$ through the 
London formula,\cite{london1937general}
\begin{equation}\label{eq:vdw-energy}
    V_{\text{London}} = -\frac{3}{2} \left(\frac{I_A I_B}{I_A+ I_B}\right) \frac{\bar{\alpha}_A\bar{\alpha}_B}{R^6},
\end{equation}
where $I_A$ is the ionization energy of system $A$, $\bar{\alpha}_A$ is the mean polarizability of $A$, and $R$ is the distance between the fragments.
This approximation neglects retardation effects appearing at long distances that modify the scaling to $R^{-7}$.\cite{CraigBook1984}
The polarizability is closely related to dipole fluctuations by the relation
\begin{equation}\label{eq:pol2dip}
    \alpha_{\gamma\gamma} = 2 \sum_{n\neq 0} \frac{\bra{\psi_0}d_\gamma\ket{\psi_n}\bra{\psi_n}d_\gamma\ket{\psi_0}}{E_n - E_0} \approx \frac{2\langle \Delta d_\gamma^2 \rangle}{I}.
\end{equation}
where $\gamma \in \{x,y,z\}$.
Inserting Eq.~\eqref{eq:pol2dip} into the London formula in Eq.~\eqref{eq:vdw-energy}, evident similarities with the dipole self-energy
\begin{equation}\label{eq:dse}
    E_{DSE} =  \frac{1}{2} \lambda^2 \langle{(\vec{\varepsilon}\cdot \Delta \vec{d})^2}\rangle.
\end{equation}
can be observed.
The London formula can also be derived from the bilinear light-matter interactions between the fragments in free space.\cite{CraigBook1984}

Cavity induced effects will be analyzed for the (H$_2$)$_2$ complex.
Three polarization directions ($\varepsilon$) are considered and the cavity frequency 
is set in resonance with the first excited state of H$_2$, $\omega=12.7$ eV.
The wave function calculations are performed with an aug-cc-pVDZ orbital basis with a bond distance of 0.74 {\AA}.
The DFT/OEP and QEDFT/OEP results are obtained using a spherical grid centered around each atom with radius 14 and spacing 0.25 bohr, and the H$_2$ 
bond distance was optimized using the KLI functional obtaining 0.73 {\AA}.

\begin{figure}[ht!]
    \centering
    \includegraphics[trim=1cm 1.2cm 1cm 0cm, clip, width=0.5\textwidth]{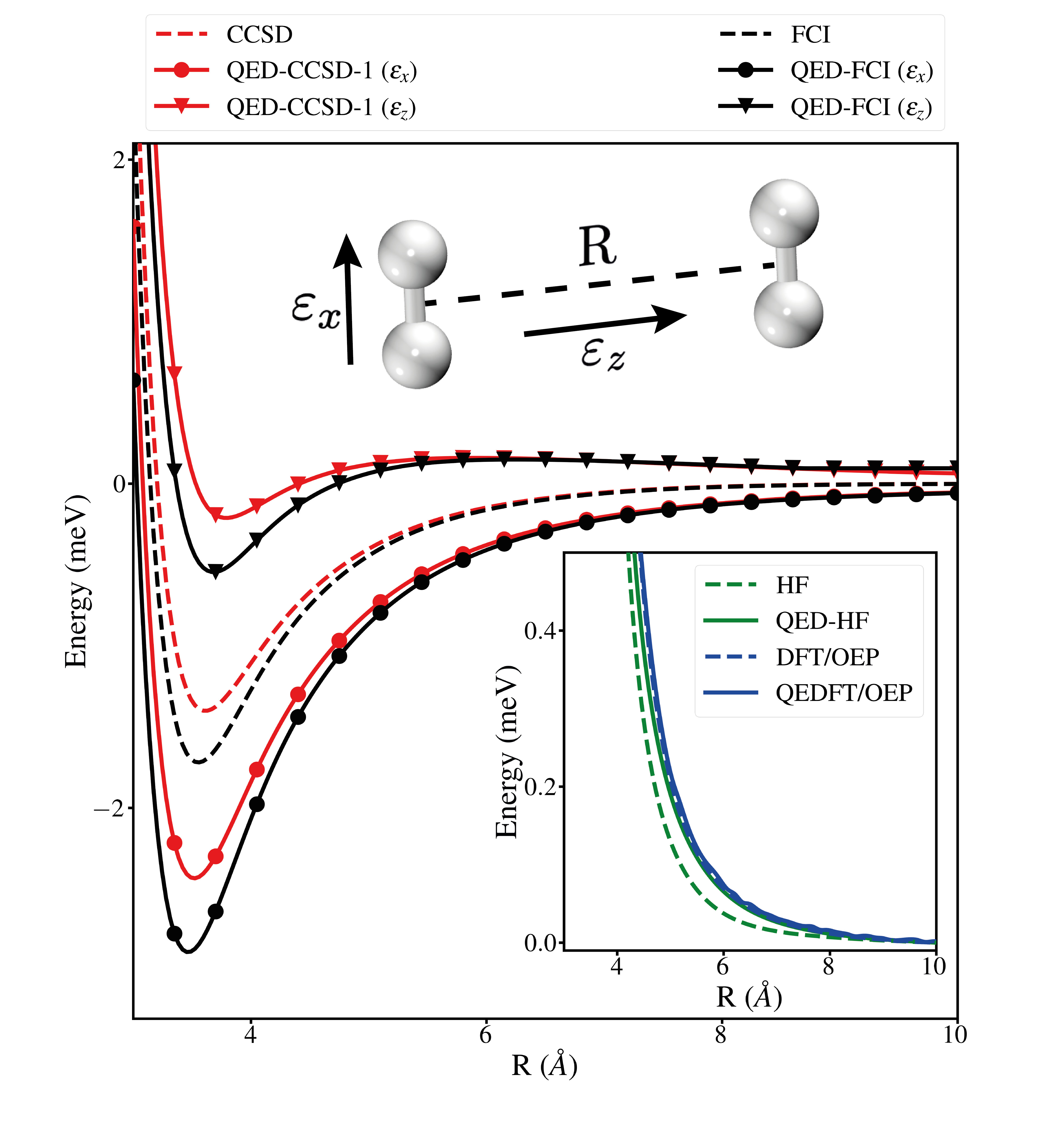}
    \caption{Potential energy curves calculated for (H$_2$)$_2$ outside (dashed line) and inside a cavity with $\lambda=0.1$ and $\omega=12.7$ eV, with different markers for the different polarizations. For each curve, the energy at 200 {\AA} has been subtracted. The inset shows the same curves for non-correlated methods with field polarization along $\varepsilon_x$.}
    \label{fig:h2h2_pes}
\end{figure}

In Figure~\ref{fig:h2h2_pes} we compare the potential energy curves calculated with CCSD/QED-CCSD-1 and FCI/QED-FCI. Field polarizations along the H$_2$ bond ($\varepsilon_x$) and chain axis ($\varepsilon_z$) are presented here. The other orthogonal polarization ($\varepsilon_y$), being qualitatively similar to $\varepsilon_x$, is shown in the Supplementary Information for completeness.

From Figure~\ref{fig:h2h2_pes}, we see that coupled cluster is able to qualitatively reproduce the reference FCI values, but it slightly underestimates the binding energy.
This is the case both outside and inside the cavity for all field polarization directions.
This behavior is well known from coupled cluster theory, and it has been extensively discussed in Ref.~\onlinecite{HelgakerBook2000}.
Including perturbative triples in QED-CC, either with CC3 or CCSD(T),\cite{koch1997cc3,urban1985ccsdt} will make the total ground state energy quantitatively more accurate compared to the reference.
Similar effects can also be observed for different molecular geometries, as shown in Figure S2 in Supplementary Information.

The Hartree-Fock method, which does not include correlation, is not able to capture the vdW interaction and the potential energy curves are repulsive, both inside and outside the cavity (see the inset in Figure~\ref{fig:h2h2_pes}).
This is in contrast to the correlated results.
The exchange-only treatment of the OEP functional provides qualitatively similar predictions compared to Hartree-Fock. 
However, in QEDFT/OEP the effect of the cavity is minimal as the DSE, bilinear and photonic contributions to the energy largely compensate each other. 

\begin{figure}[ht!]
    \centering
    \includegraphics[trim=3cm 3cm 2cm 4cm, clip, width=0.5\textwidth]{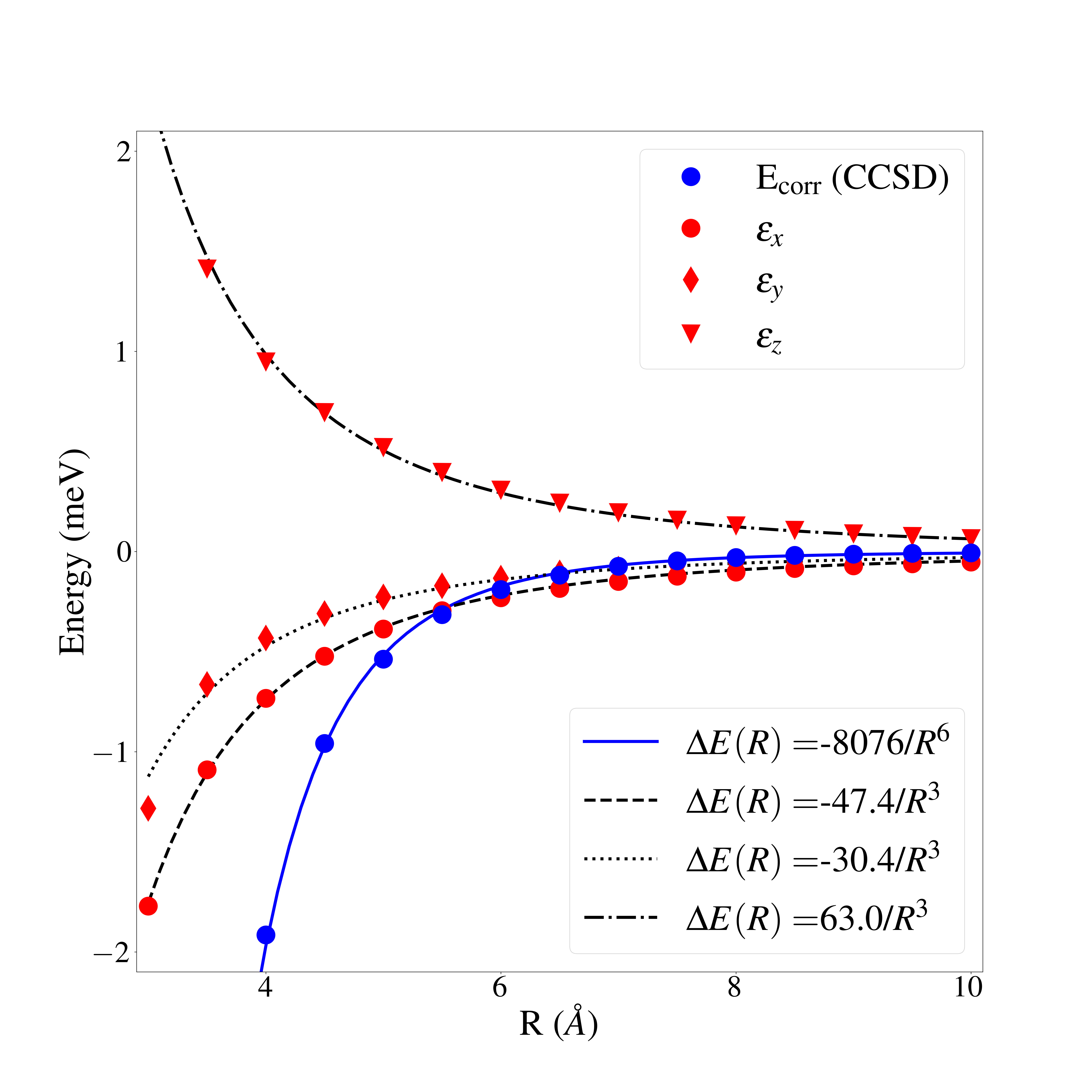}
    \caption{Correlation energy ($E_{\text{CCSD}} - E_{\text{HF}}$) and cavity induced effect on the potential energy curve of (H$_2$)$_2$ ($E_{\text{QED-CC}} - E_{\text{CC}}$) for a cavity with $\lambda=0.1$ and $\omega = 12.7$ eV. The data is fitted with a $R^{-3}$ function with $\mathcal{R}^2>0.98$ for all curves.}
    \label{fig:h2h2_induced}
\end{figure}

The cavity induced change in the potential energy curves are presented in Figure~\ref{fig:h2h2_induced}.
When the cavity polarization is perpendicular to the intermolecular bond ($\varepsilon_x$, $\varepsilon_y$), the binding energy increases.
In contrast, when the field polarization is parallel ($\varepsilon_z$), the cavity destabilizes the bond.
Interestingly, the energy difference behaves asymptotically as $R^{-3}$ for large $R$ ($\mathcal{R}^2>0.98$, black dashed lines).
This is the same distance dependence as a dipole-dipole interaction instead of an expected $R^{-6}$ behavior of the vdW dispersion.
Outside the cavity, the correlation energy (blue solid line) respects the $R^{-6}$ dependence.
Note that the dipole moments of the isolated fragments are also zero inside the cavity.

\begin{figure}[ht!]
    \centering
    \includegraphics[trim=4cm 4.5cm 4cm 1cm, clip, width=0.5\textwidth]{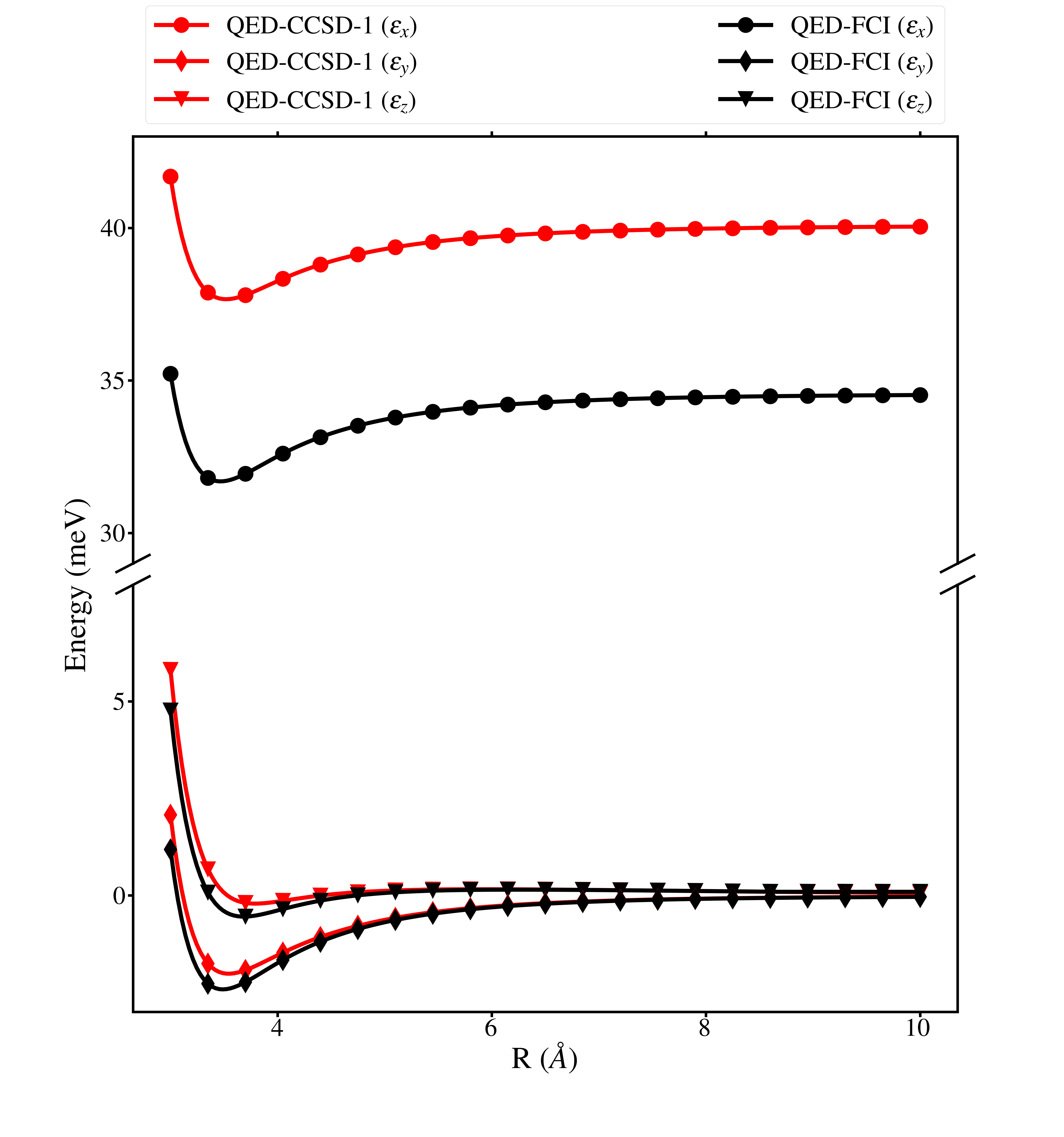}
    \caption{Potential energy curves calculated with QED-CC and QED-FCI for (H$_2$)$_2$ inside a cavity with $\lambda=0.1$ and $\omega=12.7$ eV, with different markers for the different polarizations. The energy of $\varepsilon_y$ at 200 {\AA} has been subtracted.}
    \label{fig:h2h2_nodiff}
\end{figure}

Figure~\ref{fig:h2h2_pes} and \ref{fig:h2h2_induced} show cavity induced changes in the binding energies, but they do not give any insight into the absolute stability of the system.
This is instead found from the relative energy differences (see Figure~\ref{fig:h2h2_nodiff}).
The increase in energy due to the dipole self-energy is substantially higher when the polarization is along the H$_2$ bond ($\varepsilon_x$) compared to the orthogonal field polarizations ($\varepsilon_y, \,\varepsilon_z$).
This is because the electrons are more diffuse along the bond, allowing for larger dipole fluctuations [Eq.~\eqref{eq:dse}].
The energy difference of about 40 meV induce a preferential orientation of the system along the $\varepsilon_y$ and $\varepsilon_z$ directions.

From Figure~\ref{fig:h2h2_nodiff}, we can also observe that QED-CCSD-1 is missing about 5 meV more correlation when the field polarization is along $\varepsilon_x$ compared to the other directions.
This result is confirmed by the correlation energies calculated at the minima presented in Table~\ref{tab:corr}.
\begin{table}[ht!]
    \caption{Correlation energies ($E_{\text{QED-CC/FCI}}-E_{\text{QED-HF}}$) in eV for (H$_2$)$_2$ at equilibrium geometry. The values are calculated inside and outside a cavity with coupling $\lambda=0.1$ and frequency $\omega=12.7$ eV for different field polarizations $\varepsilon$.}
    \label{tab:corr}
    \begin{ruledtabular}
    \begin{tabular}{l *{4}{D{.}{.}{1.4}}}
        Method     & \multicolumn{1}{c}{No cavity} & \multicolumn{1}{c}{$\varepsilon_x$} & \multicolumn{1}{c}{$\varepsilon_y$} & \multicolumn{1}{c}{$\varepsilon_z$} \\
        \hline
        QED-CCSD-1 & 1.9523 & 2.2400 & 2.1408 & 2.1384 \\
        QED-FCI    & 1.9526 & 2.2461 & 2.1438 & 2.1413 \\
    \end{tabular}
    \end{ruledtabular}
\end{table}

As mentioned in the discussion of Figure~\ref{fig:h2h2_pes}, the system is mainly polarizable along the H$_2$ bond ($\alpha_{xx}$) as shown in Table~\ref{tab:h2h2_pol}.
The overall reduced $\alpha$ inside the cavity indicates that the electron density becomes more localized.
The reduction is more evident along the direction of the field polarization (see columns of Table~\ref{tab:h2h2_pol}).
In this context, the variational minimization of the DSE reduces the dipole fluctuations and consequently the polarizability [see Eq. \eqref{eq:pol2dip}]. 
This observation is in line with density localization effects already discussed in Refs.~\onlinecite{FlickACSPhoton2018,schaefer2018bridging,SchaeferPNAS2019,haugland2020coupled}.
\begin{table}[ht!]
    \caption{Polarizability $\alpha_{\gamma\gamma}$ and mean polarizability $\overline{\alpha}$ for (H$_2$)$_2$ at bond distance 3.6 \AA (obtained using finite differences). All quantities are in atomic units. The cavity frequency is $\omega=12.7$ eV, and the induced effects reported for each field polarization are evaluated by subtracting the corresponding values without the cavity.}
    \label{tab:h2h2_pol}
    \begin{ruledtabular}
    \begin{tabular}{l *{4}{D{.}{.}{1.2}}}
    (H$_2$)$_2$ & \multicolumn{1}{c}{$\alpha_{xx}$} & \multicolumn{1}{c}{$\alpha_{yy}$} & \multicolumn{1}{c}{$\alpha_{zz}$} & \multicolumn{1}{c}{$\bar{\alpha}$} \\
    \hline
    CCSD            & 12.81 &  8.98 &  8.58 & 10.12 \\
    $\varepsilon_x$ & -0.18 & -0.05 & -0.06 & -0.10 \\ 
    $\varepsilon_y$ & -0.09 & -0.08 & -0.06 & -0.08 \\
    $\varepsilon_z$ & -0.10 & -0.02 & -0.09 & -0.07 \\
    \hline
    DFT/OEP         & 12.14 &  9.16 &  8.83 & 10.13 \\
    $\varepsilon_x$ & -0.11 & -0.04 & -0.04 & -0.06 \\
    $\varepsilon_y$ & -0.05 & -0.13 & -0.05 & -0.08 \\
    $\varepsilon_z$ & -0.05 & -0.05 & -0.12 & -0.07 \\
    \end{tabular}
    \end{ruledtabular}
\end{table}

The same cavity induced reduction of the polarizability is observed for an isolated H$_2$ molecule, see Table \ref{tab:h2_pol}.
This implies, using the London formula in Eq.\eqref{eq:vdw-energy}, a reduction of the vdW energy.
However, as shown in Figure~\ref{fig:h2h2_pes}, the cavity can both increase and decrease the binding energy depending on the field polarization. 
This result clearly demonstrates that the London formula, as it is usually presented, does not hold for molecules in cavities.\cite{power1982quantum,buhmann2013dispersionI}
Note that not only the polarizability is affected by the cavity, but also the ionization energy as recently shown in Ref.~\onlinecite{deprince2020cavitymodulated}.
\begin{table}[ht!]
    \caption{Polarizability $\alpha_{\gamma\gamma}$ and mean polarizability $\overline{\alpha}$ for H$_2$ (obtained using finite differences). All quantities are in atomic units. The cavity frequency is $\omega=12.7$ eV, and the induced effects reported for each field polarization are evaluated by subtracting the corresponding values without the cavity.}
    \label{tab:h2_pol}
    \begin{ruledtabular}
    \begin{tabular}{l *{4}{D{.}{.}{1.2}}}
    H$_2$ & \multicolumn{1}{c}{$\alpha_{xx}$} & \multicolumn{1}{c}{$\alpha_{yy}$} & \multicolumn{1}{c}{$\alpha_{zz}$} & \multicolumn{1}{c}{$\bar{\alpha}$} \\
    \hline
    CCSD            &  6.53 &  4.34 &  4.34 &  5.07 \\
    $\varepsilon_x$ & -0.08 & -0.03 & -0.03 & -0.05 \\
    $\varepsilon_y$ & -0.05 & -0.04 & -0.00 & -0.03 \\
    $\varepsilon_z$ & -0.05 & -0.00 & -0.04 & -0.03 \\
    \hline
    DFT/OEP         &  6.24 & 4.50 & 4.50 &  5.08   \\
    $\varepsilon_x$ & -0.06 & -0.02 & -0.02 & -0.03 \\
    $\varepsilon_y$ & -0.03 & -0.06 & -0.03 & -0.04 \\
    $\varepsilon_z$ & -0.03 & -0.03 & -0.06 & -0.04 \\
    \end{tabular}
    \end{ruledtabular}
\end{table}

\subsection{Dipole-induced dipole interactions}

In this section, the dipole-induced dipole interaction between a polar and a non-polar molecule is analyzed.
The dipole induces a charge fluctuation in the other system (induced dipole) that can form an interaction that is usually on the order of 10-100 meV.
These forces are generally stronger than the vdW dispersion.
The angle-averaged Debye formula, describing this dipole-induced dipole interaction, is given by\cite{atkins2011molecular}
\begin{equation}\label{eq:debye}
    V_{\text{Debye}} = - \frac{\vec{d}^2_A \bar\alpha_B}{R^6},
\end{equation}
where is $A$ is the polar system, and B the non-polar.
This potential has the same $R^{-6}$ behavior as the London formula in Eq.~\eqref{eq:vdw-energy}.
In symmetry adapted perturbation theory,\cite{Patkowski2020sapt} this interaction is usually known as the induction term ($E_{\text{ind}}$).

\begin{figure}[ht!]
    \centering
    \includegraphics[width=0.5\textwidth]{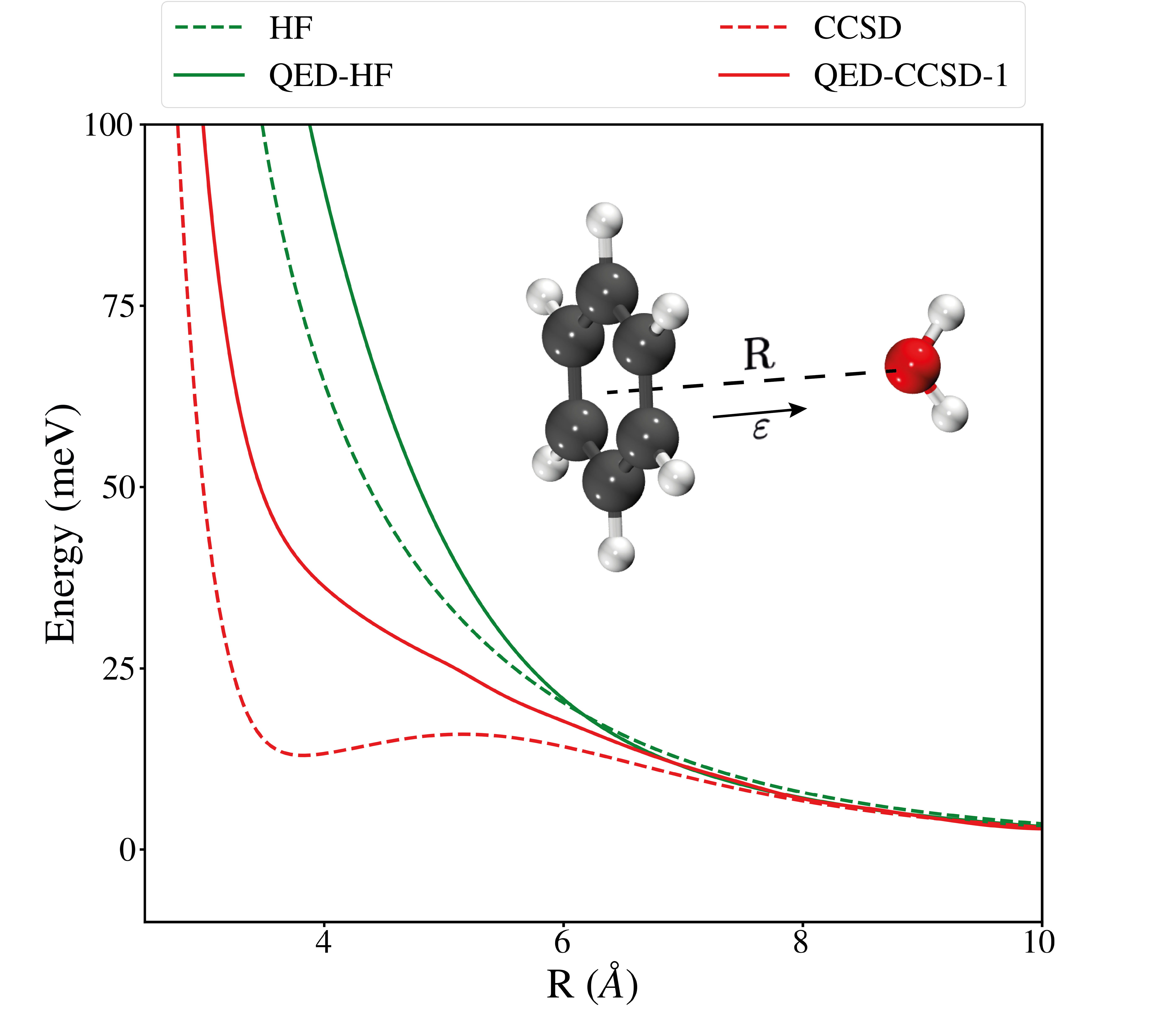}
    \caption{Potential energy curve for the Benzene-H$_2$O complex, both outside (dashed lines) and inside a cavity with cavity frequency $\omega=13.6$ eV with field polarization along the dipole. Energies are relative to the energy of the isolated fragments.}
    \label{fig:benzene-h2o}
\end{figure}

The interaction between a polar water molecule and a non-polar benzene molecule is investigated here.\cite{prakash2009benzenewater}
The complex is set in a configuration where the oxygen is pointing towards the benzene ring, as shown in the inset of Figure~\ref{fig:benzene-h2o}. 
In this geometry, the interaction is dominated by the dipole-induced dipole forces and hydrogen bonding is minimal.
The structures of the separate fragments are optimized using DFT/B3LYP with a 6-31+G** basis set.

In Figure~\ref{fig:benzene-h2o} we show the potential energy curves calculated using HF, CCSD, QED-HF, and QED-CCSD-1 with a 6-31+G* basis set.
The field polarization is set along the dipole (see Figure~\ref{fig:benzene-h2o}).
We first note that HF and QED-HF are repulsive outside and inside the cavity.
For CCSD, the weakly interacting complex has a meta-stable state with a barrier of 3 meV.
Inside the cavity, instead, the QED-CCSD-1 potential energy curve becomes repulsive.
The cavity induced effect is here much larger (26 meV) than the observed effects for the non-polar H$_2$ vdW interaction.

From the Debye formula in Eq.~\eqref{eq:debye}, we see that the strength of the induction interaction is related to the dipole of the polar fragment and the polarizability of the non-polar one.
In Table \ref{tab:benzene-h2o_pol} the dipole moment of H$_2$O and the polarizability of benzene, outside and inside the cavity, are provided.
Inside the cavity we see a small decrease (0.003 D) of the permanent dipole, but a sizable reduction (7.60 a.u.) of the benzene polarizability.
A possible reason for the large cavity effect on $\alpha$ could be connected to the aromaticity of benzene and its diffuse electrons.
However, this will require further investigations.\cite{Lazzeretti2004aromaticity}
\begin{table}[ht!]
    \caption{Dipole $d$ for water and polarizability $\alpha$ parallel and perpendicular to $d$ for Benzene. Calculated outside and inside a cavity with field polarization along $d$ and frequency $\omega=13.6$ eV.}
    \label{tab:benzene-h2o_pol}
    \begin{ruledtabular}
    \begin{tabular}{lcccc}
    Method & $\alpha_{\perp}$ & $\alpha_{\parallel}$ & $\bar{\alpha}$ & $d$ [D] \\
    \hline
    CCSD       & 78.55 & 40.83 & 65.97 & 2.304  \\
    QED-CCSD-1 & 77.60 & 33.23 & 62.81 & 2.301  \\
    \end{tabular}
    \end{ruledtabular}
\end{table}

Substituting the values in Table~\ref{tab:benzene-h2o_pol} into Eq.~\eqref{eq:debye}, and comparing the size of the interaction outside and inside the cavity, we estimate a destabilization of 0.5 meV at $R=3.8$ {\AA}.
This is much smaller than the corresponding destabilization observed in Figure~\ref{fig:benzene-h2o} of about (26 meV).
This indicates that, as in the case of vdW, the purely electrostatic interaction model is not sufficient to reproduce the effect.

\subsection{Hydrogen bonds}

Hydrogen bonding arises when a hydrogen atom forms a bridge between two electronegative species.\cite{IUPAC2011hydrogenbond}
A typical hydrogen bond energy is about 100-300 meV and is one of the stronger intermolecular interactions.
The electrostatic dipole-dipole energy is given by\cite{CraigBook1984}
\begin{equation}\label{eq:dipole}
    V_{\text{dipole}} = \frac{1}{R^{3}} \left(\vec{d}_A \cdot \vec{d}_B - \frac{3 (\vec{d}_A \cdot \vec{R})(\vec{d}_B \cdot \vec{R})}{R^2} \right)
\end{equation}
and represents the leading term to this interaction.
However, it is now widely accepted that this kind of interaction is also characterized by a sizable charge transfer component, which contributes with a relatively large percentage (about 2-3 eV per transferred electron) to the binding energy.\cite{CappellettiAccChemRes2012,Ronca2014waterdimer}

Investigating hydrogen bonding is crucial to understand cavity induced effects on the physical and chemical properties of solvents. 
In this section, we investigate the water dimer as a simple model for liquid water.
For HF and CC calculations, we use an aug-cc-pVDZ basis set.
The OEP calculations are performed using a spherical grid centered around each atom with radius 12 and spacing 0.28 bohr.
The MP2/aug-cc-pVQZ equilibrium geometry is obtained from Ref.~\onlinecite{Ronca2014waterdimer}.
The field polarization is set along the O-O direction, and the cavity frequency is $\omega=7.86$ eV.
\begin{figure}[ht!]
    \centering
    \includegraphics[trim=0cm 0.5cm 1.5cm 0cm, clip, width=0.5\textwidth]{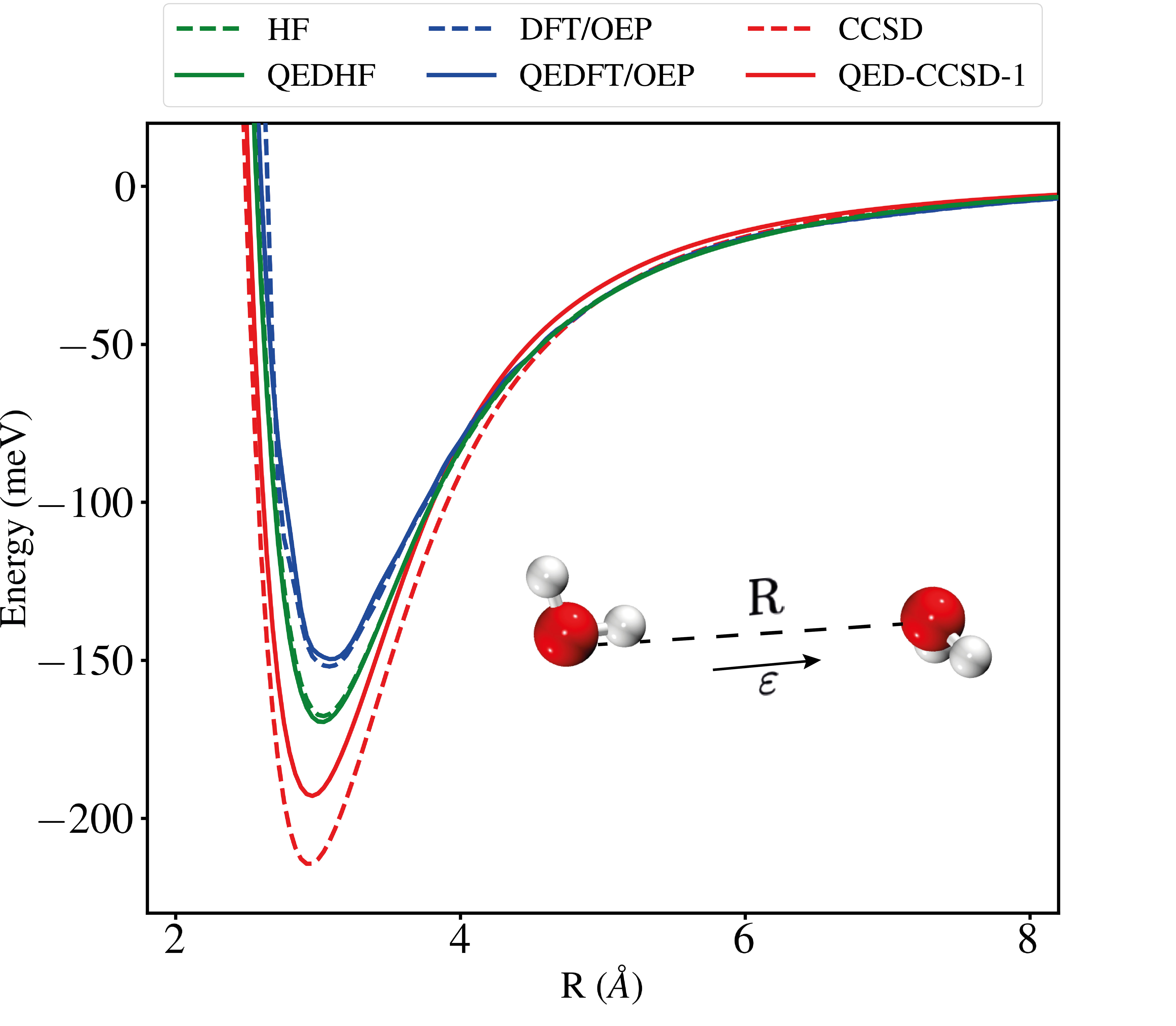}
    \caption{Potential energy curves of (H$_2$O)$_2$ for different distances $R$ between the oxygen atoms, outside (dashed) and inside a cavity (solid). The cavity polarization is along $R$ with frequency $\omega=7.86$ eV. The structures of the water fragments are fixed.}
    \label{fig:h2o-stretch}
\end{figure}

In Figure~\ref{fig:h2o-stretch} we present the potential energy curve of the water dimer as a function of the O-O distance ($R$).
In this case, the non-correlated methods (HF and DFT/OEP) capture the bond.
Outside the cavity, DFT/OEP gives a slightly smaller binding energy than HF, and they both underestimate the energy compared to CCSD ($214$ meV).
Inside the cavity, QED-CCSD-1 predicts a weaker hydrogen bond.
The non-correlated methods (QED-HF and QEDFT/OEP) do not capture this cavity induced destabilization of about $22$ meV (10\%), which indicates that the effect is due to correlation.

The electrostatic contribution to the binding energy can be estimated using the dipole moments shown in Table~\ref{tab:h2odip} and Eq.~\eqref{eq:dipole}.
The perturbative treatment predicts a cavity induced stabilization of about 0.2 meV, in contrast to the destabilization shown in Figure~\ref{fig:h2o-stretch}.
From this observation, it is clear that cavity induced effects are not due to shifts in the multipoles (dipole, etc.) using a purely electrostatic picture. 

\begin{table}[ht!]
    \caption{Dipole moments of isolated water fragments $A$ and $B$, outside and inside the cavity. The field polarization is along the O-O direction and $\omega=7.86$ eV. In the hydrogen bond, $A$ is the donor while $B$ is the acceptor.}
    \label{tab:h2odip}
    \begin{ruledtabular}
    \begin{tabular}{lcc}
    Method & $d_A$ [D] & $d_B$ [D] \\
    \hline
    CCSD       & 1.853 & 1.853 \\
    QED-CCSD-1 & 1.860 & 1.850 \\
    DFT/OEP    & 2.014 & 2.014 \\
    QEDFT/OEP  & 2.025 & 2.021 \\
    \end{tabular}
    \end{ruledtabular}
\end{table}

\begin{figure}[ht!]
    \centering
    \includegraphics[width=0.5\textwidth]{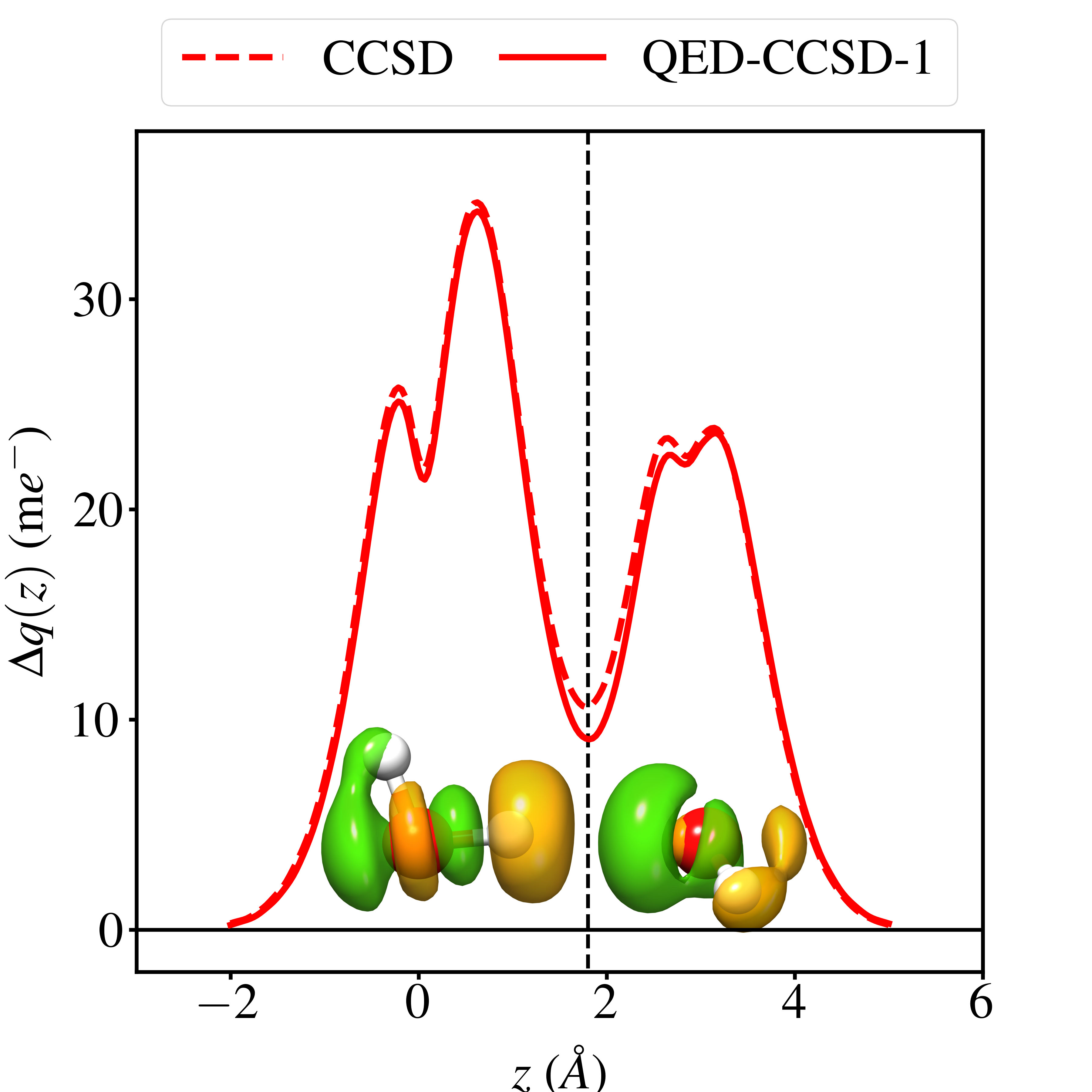}
    \caption{CCSD and QED-CCSD-1 bond formation charge displacement analysis of (H$_2$O)$_2$ inside (dashed) and outside (solid) a cavity. The fragments are separated by 3.0 {\AA}. The cavity polarization is along $z$.}
    \label{fig:h2o_dim_dens}
\end{figure}

As discussed above, hydrogen bondings are characterized by a relatively large charge transfer contribution.
To investigate this aspect, we perform a charge displacement (CD) analysis\cite{BelpassiJAmChemSoc2008, CappellettiAccChemRes2012} in Figure~\ref{fig:h2o_dim_dens}.
The charge displacement function ($\Delta q(z)$) is defined from the density difference $\Delta\rho = \rho_{Dimer}-(\rho_{(\mathrm{H}_2\mathrm{O})_A}+\rho_{(\mathrm{H}_2\mathrm{O})_B})$ as
\begin{equation}
    \Delta q(z) = \int_{-\infty}^{\infty}\int_{-\infty}^{\infty}\int_{-\infty}^{z} \Delta\rho(x,y,z') \, \text{d}x\,\text{d}y\,\text{d}z'.
\end{equation}
This function quantitatively describes the amount of charge that has been moved along the z-direction.
In particular, when $\Delta q$ is positive, electrons are moved to the left, and they change direction when it is negative.
For the water dimer outside the cavity, electrons are moved from the donor molecule (right) to the acceptor (left).
A net charge transfer of about 10 m$e^-$ can be observed in the middle (dashed line) of the hydrogen bond. 
By placing the system inside the cavity, the number of transferred electrons is reduced by about 2 m$e^-$. 
This is in line with the weaker hydrogen bond observed in the potential energy curve and, following Ref.~\onlinecite{CappellettiAccChemRes2012}, it corresponds to a decrease of about 4-6 meV in the binding energy.

\begin{figure}[ht!]
    \centering
    \includegraphics[width=0.5\textwidth]{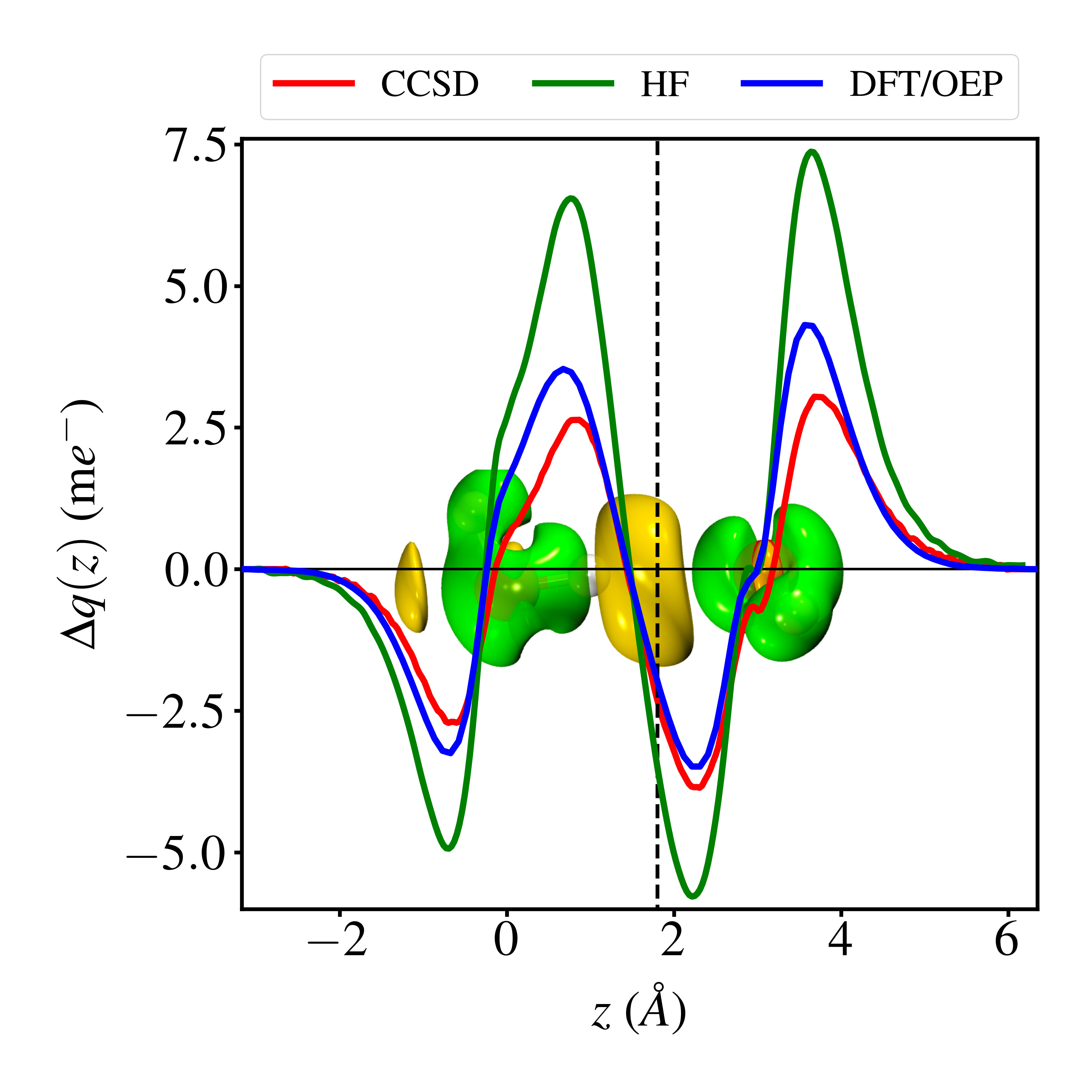}
    \caption{Charge displacement analysis of the ground state cavity induced density difference of (H$_2$O)$_2$. The fragments are separated by 3.0 {\AA}. The cavity polarization is along $z$.}
    \label{fig:h2o_dim_dens_conf}
\end{figure}

In Figure~\ref{fig:h2o_dim_dens_conf}, the CD analysis has been applied to the ground state cavity induced density difference ($\Delta\rho= \rho^{\mathrm{GS}}_{\mathrm{cav}}-\rho^{\mathrm{GS}}_{\mathrm{nocav}}$) calculated using HF, CCSD, and DFT/OEP. 
In this case, all methodologies show a qualitatively similar behaviour and predict a reduction of charge transfer from donor to acceptor of about 2-3 m$e^-$ (extracted at the same position as in Figure~\ref{fig:h2o_dim_dens}). The overall effect is a charge localization on the water fragments. The reduction of charger-transfer for small distances is consistent with the observations of Ref.~\onlinecite{SchaeferPNAS2019} and originates mainly from the DSE.
From a more quantitative analysis, we see that Hartree-Fock overestimates the cavity induced charge transfer compared to coupled cluster. On the other hand, the DFT/OEP curve is closer to the correlated result.
In general, DFT is better equipped to describe the density than Hartree-Fock.\cite{kummel2008orbital}

\subsection{Cavity induced non-additive properties}

Even when molecules are far away from each other inside a cavity, they still interact.\cite{SchaeferPNAS2019}
This results in non-additive properties of the dissociated system, \textit{e.g.} the energy of the dissociated complex will not be equal to the sum of the energies of the individual fragments.
In this section, we investigate these cavity induced non size-extensive properties of the QED Hamiltonian and their implications.

The main non-additive effect can be attributed to the DSE, which rewritten in terms of contributions from the individual fragments takes the form
\begin{align}
    E_{\text{DSE}}^{\mathrm{complex}} &= \sum_n^{N_f} E_{\text{DSE}}^{(n)} + \lambda^2 \sum_{n<m}^{N_f} \langle (\vec{\varepsilon}\cdot \Delta \vec{d}^{(m)})(\vec{\varepsilon}\cdot \Delta \vec{d}^{(n)}) \rangle,
    \label{eq:sizeextensive}
\end{align}
where $N_f$ is the number of fragments, $\Delta \vec{d} = \sum_n \Delta \vec{d}^{(n)}$ and $E^{(n)}_{\text{DSE}}$ is the DSE for fragment $n$.
The last term in Eq.~\eqref{eq:sizeextensive} is the one non-additive part.
For identical fragments, the expression simplifies to
\begin{equation} \label{eq:sizeext-quad}
    \Delta \mathcal{E}_{\text{DSE}} = \frac{\lambda^2}{2}\langle (\vec{\varepsilon}\cdot \Delta \vec{d}^{(1)}) (\vec{\varepsilon}\cdot \Delta \vec{d}^{(2)}) \rangle (N_{f}^2 - N_{f}),
\end{equation}
showing explicitly that the effect is quadratic with the number of fragments.
In Eq. \eqref{eq:sizeext-quad}, $\Delta \vec{d}^{(1)}$ and $\Delta \vec{d}^{(2)}$ are the dipole fluctuations of two equivalent fragments.
The bilinear term will also contribute,
\begin{equation} \label{eq:size_ext_bi}
    E_{\text{BI}}^{\text{complex}} = \lambda \sqrt{\frac{\omega}{2}} \sum_n^{N_f} \langle (b+b^\dagger) (\vec{\varepsilon}\cdot \Delta \vec{d}^{(n)}) \rangle,
\end{equation}
although we expect the contribution to be smaller.

\begin{figure}[ht!]
    \centering
    \includegraphics[trim=0cm 0cm 1cm 2cm, clip, width=0.5\textwidth]{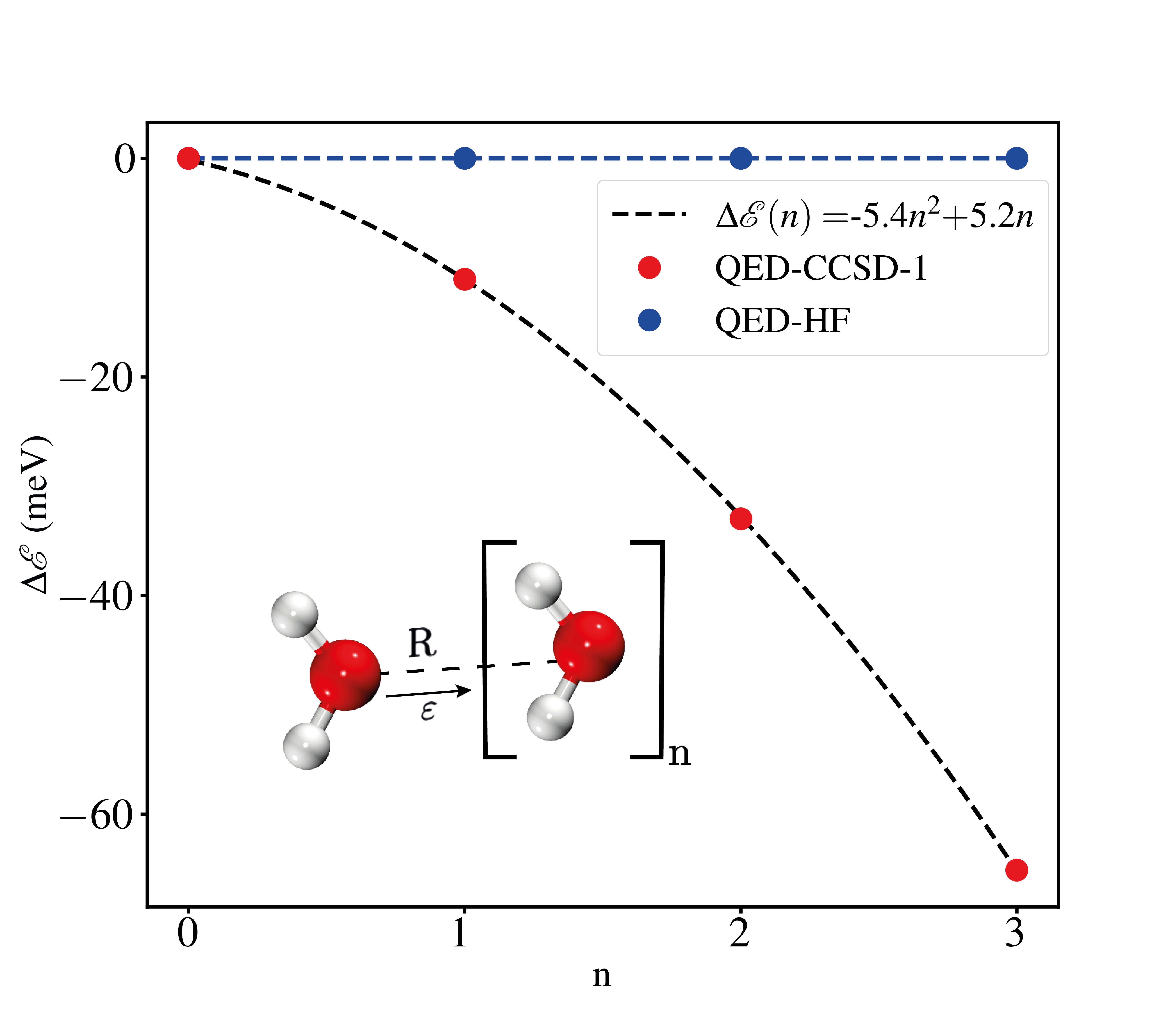}
    \caption{The non-additive part of the total energy, $\Delta \mathcal{E} = \mathrm{E}_{(\mathrm{H_2O})_{n+1}} - (n+1)\mathrm{E}_{\mathrm{H_2O}}$ of (H$_2$O)$_{n+1}$. Fragments are separated by 200 {\AA} and placed inside a cavity with $\omega=7.86$ eV. A 6-31G basis set is used. In black dashed lines is a polynomial fitting with $\mathcal{R}^2>0.999$.}
    \label{fig:size_ext}
\end{figure}

In Figure~\ref{fig:size_ext}, we analyze the non-additive part of the total energy ($\Delta\mathcal{E}=E^{\mathrm{complex}}-\sum_n E^{(n)}$) for water molecules separated by 200 {\AA} (see inset of Figure~\ref{fig:size_ext}).
The properties of the QED Hamiltonian makes the total energy non-additive, and the QED-CC approach is able to capture this behavior (black dashed line).
In the limit $\lambda\rightarrow 0$, the QED-CC approach is size-extensive as for standard coupled cluster theory for electrons.
On the other hand, QED-HF always has size-extensive solutions and fails to describe the non-additive nature of the cavity interaction (blue dashed lines).

We observe a quadratic scaling of the non-additive part with respect to the number of fragments, as expected from Eq.~\eqref{eq:sizeext-quad}.
The fit of the data (dashed black line) highlight also a small deviation (0.2) in the linear term, likely due to the bilinear correction described in Eq.~\eqref{eq:size_ext_bi}.
This result implies that the collective ground state interaction between many molecules could be observable experimentally, if enough molecules are considered.
We point out that this effect is qualitatively not limited by the dipole approximation, since we consider a stretching of the coordinates in a direction that is parallel to the field polarization (perpendicular to the wave vector $\vec{k}$).
This non-additive effect could be very interesting in solvation environments, where a solute can interact with a large number of solvent molecules through the field.  

\section{Conclusions} \label{sec:conclusions}

Leveraging available QED \textit{ab initio} methodologies, we investigated cavity induced effects on intermolecular interactions.
Four types of interactions have been studied: van der Waals forces, induction interactions, hydrogen bonding, and cavity mediated long-range interactions.
In all cases, correlation is deemed crucial to describe the system.
Several effects of the cavity seem rather counter-intuitive at first glance.
The van der Waals forces between two non-polar molecules behaves as $R^{-6}$.
However, inside the cavity, an additional $R^{-3}$ component originates from correlation.
For the induction interaction and hydrogen bonding, the cavity induced effect is larger than for the van der Waals.
In both cases, a destabilization of the binding energy and a reduction of polarizability is observed.
We show that, inside the cavity, molecules remain permanently correlated at arbitrary distances.

Our results suggest that cavity fields can be used to modify the ground state interactions in intermolecular systems.
This opens possibilities for novel applications in several fields, going from the control of solvent assisted reactions to modifying higher-order structures in biological macromolecules.
We think that these results will be crucial to understand future experiments in this direction.

\begin{acknowledgments}
T.S.H. and H.K. acknowledge computing resources through UNINETT Sigma2 - the National Infrastructure for High Performance Computing and Data Storage in Norway, through project number NN2962k. T.S.H. and H.K. acknowledge funding from the Marie Sk{\l}odowska-Curie European Training Network “COSINE - COmputational Spectroscopy In Natural sciences and Engineering”, Grant Agreement No. 765739, the Research Council of Norway through FRINATEK projects 263110 and 275506. 
C.S. and A.R. acknowledge support of the RouTe Project (13N14839), financed by the Federal Ministry of Education and Research (Bundesministerium f{\"u}r Bildung und Forschung (BMBF)), the European Research Council (ERC-2015-AdG694097), the Cluster of Excellence ``Advanced Imaging of Matter" (AIM) and Grupos Consolidados (IT1249-19). The Flatiron Institute is a division of the Simons Foundation.
\end{acknowledgments}

\section*{Data Availability Statement}
The data that supports the findings of this study are available within the article and its supplementary material.

\bibliography{bib}

\end{document}